\documentstyle[sprocl,psfig]{article}

\def\Journal#1#2#3#4{{#1} {\bf #2}, #3 (#4)}

\def\NPB{{\em Nucl. Phys.} B}
\def\PLB{{\em Phys. Lett.}  B}
\def\PRL{\em Phys. Rev. Lett.}
\def\PR{\em Phys. Rev.}
\def\PRD{{\em Phys. Rev.} D}
\def\ZPC{{\em Z. Phys.} C}
\def\PRC{{\em Phys. Rev.} C}
\def\NPA{{\em Nucl. Phys.} A}
\def\JMP{{\em Int. J. Mod. Phys.} E}
\def\RPP{\em Rep. Progr. Phys.}
\def\PP{\em Phys. Rep.}

\def\be{\begin{equation}}
\def\ee{\end{equation}}
\def\bea{\begin{eqnarray}}
\def\eea{\end{eqnarray}}

\begin{document}

\title{Nucleon Electromagnetic Form Factors in a chiral constituent-quark model}

\author{P.~Demetriou}

\address{Insitute of Nuclear Physics, N.~C.~S.~R. ``Demokritos'',\\
Aghia Paraskevi GR-156 10, Athens, Greece\\e-mail: vivian@mail.demokritos.gr} 

\author{S.~Boffi, M.~Radici and R.~F.~Wagenbrunn}

\address{Dipartimento di Fisica Nucleare e Teorica, \\
Universit\`a di Pavia and INFN, Pavia 27100, Italy}

\maketitle\abstracts{The electromagnetic form factors of the nucleon are calculated
in an extended chiral constituent-quark model where the effective interaction is
described by the exchange of pseudoscalar, vector, and scalar mesons. Two-body
current-density operators, constructed consistently with the extended model
Hamiltonian in order to preserve gauge invariance and current conservation, are found
to give a significant contribution to the nucleon magnetic form factors and improve the
estimates of the nucleon magnetic moments.}

\section{Introduction}

Constituent quark models (CQM) have been widely used to describe the spectroscopic 
properties of hadrons and have been rather successful in
reproducing the gross features of hadron spectra within a 
nonrelativistic~\cite{DeRujula,IK,Giannini,Karl} and relativistic 
framework~\cite{Warns,Capstick,Cardarelli,Cotanch}.
 In all these models the effective
interaction between the valence quarks is described  by the one-gluon-exchange diagram and is
identified with the hyperfine-like part of its nonrelativistic reduction.
Various hybrid models have also been constructed including meson exchanges in
addition to sizeable contributions coming from gluon exchanges~
\cite{Obukhovsky,BHF,valcarce,fabre,Shen}.

Despite the overall success, none of these models has been able to explain 
the correct level orderings in light- and strange baryon spectra nor the flavour or spin content
of the nucleon. This is mainly due to the inadequacy of interactions that do not 
take into account the implications of the spontaneous breaking of
chiral symmetry (SB$\chi$S). As a consequence of SB$\chi$S, quarks acquire their dynamical
masses  related to $<q\bar q>$ condensates and Goldstone bosons appear which couple
directly to the constituent quarks. Thus beyond the scale of SB$\chi$S the effective degrees of
freedom are constituent quarks and Goldstone-boson fields and baryons can be considered as
systems of three constituent quarks that interact by Goldstone-boson exchange (GBE) and are
subject to confinement~\cite{Manohar,gloz1}. The Goldstone bosons manifest themselves in the
octet of pseudoscalar mesons ($\pi,\,K,\,\eta$).

In view of these considerations, a GBE CQM has been
proposed~\cite{gloz2,gloz3}  based on a semirelativistic Hamiltonian where the
dynamical part consists of a linear confinement potential and a chiral potential containing 
the spin-spin components of the pseudoscalar meson exchange
interaction.
The model is able to reproduce the correct level orderings of positive- and negative-parity
excitations providing hence a unified description not only of the
Nucleon and Delta spectra but also of all strange baryon spectra.

A further, stringent test of the model is to investigate its validity 
with regard to other observables. Such an attempt, where the three-Q wavefunctions obtained 
from the pseudoscalar-exchange version of the semirelativistic GBE
CQM  were used to calculate the elastic electromagnetic form factors of the 
nucleon~\cite{boffi1}, has shown that the two-body current
operator constructed consistently with the model Hamiltonian gives zero contributions in this
case. Furthermore, the semi-relativistic 
one-body charge- and current-density operators~\cite{Capuzzi} underpredict the charge radii and 
the magnetic moments of the nucleon. However, the two-body currents arising in the
 pseudoscalar-exchange version of the model are due solely to the spin-spin
component of the pseudoscalar-exchange
interaction, whereas we would expect that the current operators obtained from the full
pseudoscalar meson-exchange interaction including the tensor component would give 
non-zero contributions to the form factors altering the picture obtained in ref.~\cite{boffi1}. 

The  importance of  two-body currents for the electromagnetic properties of
baryons is therefore still not well understood and it is the aim
of this contribution
to gain further insight into the  different Q-Q interactions and the relative
contributions of the exchange-currents they give rise to within
an extended GBE CQM, where tensor forces have been taken into account. 

\section{Extended GBE CQM}

In the extended GBE CQM~\cite{gloz4} the three-quark Hamiltonian is  
\be 
H_0 = \sum_{i=1}^3 \sqrt{{\vec p}^{\,2}_i + m_i^{2}} + \sum_{i<1}^3 V_{ij}
\label{eq:kinetic}
\ee
with $m_i$ the masses and ${\vec p}_i$ the three-momenta of the constituent
quarks. This form ensures that the average quark velocity be lower than the
light velocity, a requirement that is not fulfilled by nonrelativistic models. The dynamical 
part consists of a Q-Q interaction 
\be
V_{ij} = V_{conf} + V_{\chi},
\ee
with a central confining interaction $V_{conf}$ and the chiral
interaction $V_{\chi}$.
The latter contains spin-spin, tensor, and central forces coming
from pseudoscalar meson exchanges as well as from vector and scalar meson exchanges, as a 
representation for multiple GBE (see also ref.~\cite{gloz4} in these proceedings for further
details). The baryon spectra of the extended GBE CQM
are very similar to the ones obtained already in the GBE CQM
of refs.~\cite{gloz2,gloz3}, where only the spin-spin component from
the pseudoscalar meson exchanges have been employed.

\section{The charge-current operators}

The relativistic form of the kinetic energy does not permit the use of the
traditional one-body current density operator; so in order to be consistent with the model
Hamiltonian the gauge invariant charge-current density operator is derived within a
functional derivative formalism~\cite{Capuzzi}. It contains both one- and two-body terms. The
one-body contribution includes the charge, the convective- and the spin-current
operators. Their matrix elements between free particle states for a particle of 
charge $e$ and mass $m$  have been derived in momentum space~\cite{Capuzzi} and with 
respect to the usual nonrelativistic expressions, only the spatial components of the 
charge-current density operator are
affected, while the time component is simply given by the charge density. 

The two-body current operator can be derived directly from the continuity
equation consistently with the model Hamiltonian
of the extended GBE CQM.
In momentum space the continuity equation reads

\bea
{\vec q} \cdot {\vec J}_{[2]} = \left[ \bar J^0_{[1]},\bar V \right] 
\label{eq:continuity}
\eea

\noindent
with $\bar J^0_{[1]}, \bar V$ the Fourier transforms of the one-body charge-density and 
Q-Q potential 
$V_{ij}$ of the previous section, respectively. Due to the flavor-dependence of
$J^0_{[1]}$ it turns out that the only non-vanishing exchange currents arise from $\pi$-, 
$K$-, $\rho$- and 
$K^*$-exchange. If we restrict ourselves to the non-strange baryon sector, then
we have contributions only from the exchange of pions and rho-mesons and 
due to their isospin structure the exchange currents we finally obtain from
(\ref{eq:continuity}) are the well-known pion(rho)-pair ($\pi(\rho) q\bar q$) currents and
pion(rho)-in-flight ($\gamma \pi(\rho) \pi(\rho)$) currents

\bea 
{\vec J}_{\pi q\bar q}({\vec k}_i, {\vec k}_j) = ie\,\frac{g_{\pi}^2}{4m_i m_j}\Big[
\frac{{\vec\sigma}_i\cdot {\vec k}_i}{({\vec k}_i^2+\mu_{\pi}^2)}\,
\Big(\frac{\Lambda_{\pi}^2-\mu_{\pi}^2}{{\vec k}_i^2+\Lambda_{\pi}^2}\Big)^2\,
{\vec\sigma}_j-\, ( i \leftrightarrow j )\Big]\,({\vec \tau}_i \times {\vec \tau}_j)_z 
\eea

\bea 
{\vec J}_{\gamma \pi\pi}({\vec k}_i, {\vec k}_j) &=& ie\,\frac{g_{\pi}^2}{4m_i m_j}
\frac{{\vec\sigma}_i\cdot {\vec k}_i\,{\vec\sigma}_j\cdot {\vec k}_j}
{({\vec k}_i^2+\mu_{\pi}^2)({\vec k}_j^2+\mu_{\pi}^2)}\,({\vec k}_i - {\vec k}_j)\,
\frac{(\Lambda_{\pi}^2-\mu_{\pi}^2)^2}{({\vec k}_i^2+\Lambda_{\pi}^2)({\vec k}_j^2+
\Lambda_{\pi}^2)}\, \nonumber \\
 &{}& \qquad \times\,\Big(1 + \frac{{\vec k}_i^2+\mu_{\pi}^2}{{\vec k}_j^2+
\Lambda_{\pi}^2}+\frac{{\vec k}_j^2+\mu_{\pi}^2}{{\vec k}_i^2+
\Lambda_{\pi}^2}\Big)\,({\vec \tau}_i \times {\vec \tau}_j)_z 
\eea

\bea 
{\vec J}_{\rho q\bar q}({\vec k}_i, {\vec k}_j) &=& ie\,\frac{(g_{\rho}^V+g_{\rho}^T)^2}
{4m_i m_j}\Big[
\frac{{\vec\sigma}_i \times ({\vec\sigma}_j \times {\vec k}_j)}{({\vec k}_j^2+\mu_{\pi}^2)}\,
\Big(\frac{\Lambda_{\rho}^2-\mu_{\rho}^2}{{\vec k}_j^2+\Lambda_{\rho}^2}\Big)^2-
\, ( i \leftrightarrow j )\Big] \nonumber \\
 &{}& \qquad\,\times\,({\vec \tau}_i \times {\vec \tau}_j)_z 
\eea

\bea 
{\vec J}_{\gamma \rho\rho}({\vec k}_i, {\vec k}_j) &=& ie\,\Big[(g_{\rho}^V)^2+
\frac{(g_{\rho}^V+g_{\rho}^T)^2}{4m_i m_j}\,({\vec\sigma}_i\times {\vec
k}_i)\cdot({\vec\sigma}_j\times {\vec k}_j)\Big]
\frac{({\vec k}_i - {\vec k}_j)}
{({\vec k}_i^2+\mu_{\rho}^2)({\vec k}_j^2+\mu_{\rho}^2)}\, \nonumber \\
&{}&  \times\,
\frac{(\Lambda_{\rho}^2-\mu_{\rho}^2)^2}{({\vec k}_i^2+\Lambda_{\rho}^2)({\vec k}_j^2+
\Lambda_{\rho}^2)}\, \Big(1 + \frac{{\vec k}_i^2+\mu_{\rho}^2}{{\vec k}_j^2+
\Lambda_{\rho}^2}+\frac{{\vec k}_j^2+\mu_{\rho}^2}{{\vec k}_i^2+
\Lambda_{\rho}^2}\Big)\,({\vec \tau}_i \times {\vec \tau}_j)_z 
\eea

\noindent
where ${\vec q} = {\vec k}_i + {\vec k}_j$, $\mu_\pi$ and $\mu_\rho$ are
the meson masses, $g_\pi$, $g_{\rho}^V$, and  $g_{\rho}^T$ are the pion-quark,
the $\rho$-quark vector, and the $\rho$-quark tensor coupling constants,
and $\Lambda_\rho$, $\Lambda_\pi$ are cut-off parameters which are connected with
extended meson-quark vertices. For all parameters the values quoted
in ref.~\cite{gloz4} have been used.

\section{Results}

The calculation of the elastic e.m. form factors involves the calculation of the matrix
elements of the charge- and current-density operators presented in the previous section.
The electric form factor consists of contributions from the one-body charge-density
operator $J^0_{[1]}$ whereas the magnetic form factor consists of contributions
from the one-body and two-body current-density operators. The full magnetic form factor
 can be written as the sum of the individual 
 contributions 

\be
G_M (Q^2) = G_M^{[1]}(Q^2) + G_M^{\pi q\bar q}(Q^2) + 
G_M^{\gamma \pi\pi}(Q^2) + G_M^{\rho q\bar q} + G_M^{\gamma \rho\rho}(Q^2)
\ee

The calculations were performed without introducing any additional parameters and the
results for the electric and magnetic form factors are plotted in figs.~1 and~2. There is no 
modification in the calculated 
electric form factor since two-body effects do not appear in the charge-density 
operator. However, contrary to the case of a pseudoscalar meson-exchange
interaction restricted to its  spin-spin component only
where the contribution from two-body currents
was found to be zero~\cite{boffi1}, we find that currents
arising from the full pion- and rho-exchange interaction give a sizeable contribution to
the magnetic form factors which gradually decreases with increasing $Q^2$. 
Contributions from the pion pair-currents and pionic currents tend to cancel each other
(the same applies to the rho-exchange currents) but the overall effect is the
enhancement of the nucleon form factors. At $Q^2$ = 0
in particular, the contribution from two-body currents to the magnetic moment 
$\mu_{N} = G_M (Q^2=0)$ of the proton and neutron is significant, giving a much better 
agreement with the
experimental values compared to the results with one-body currents only 
as shown in Table~\ref{tab:1}.

\begin{table}[h]
\caption{Contributions to the magnetic moments of the proton and neutron from different
currents.\label{tab:1}}
\vspace{0.2cm}
\begin{center}
\footnotesize
\begin{tabular}{|c|c|c|c|c|c|c|}
\hline
N & $\mu_{[1]}$ & $\mu_{\pi q\bar q}$ & $\mu_{\gamma \pi\pi}$ & $\mu_{\rho q\bar q}$ &
 $\mu_{\gamma \rho\rho}$ & $\mu^N$ \\ \hline
 p& 1.516&-0.126&0.735&-0.109&0.202&2.218 \\ \hline
 n&-0.993&0.126&-0.735&0.109&-0.202&-1.695 \\ 
\hline
\end{tabular}
\end{center}
\end{table}

Despite the improvement at $Q^2$ = 0, the  electric and magnetic form
factors overestimate the experimental form factors at $Q^2 \ne 0$ and hence lead to 
an underestimation of the nucleon charge radii. This is a common feature of all CQM's
and reflects the fact that constituent quarks are assumed to be pointlike particles. 
Any effects resulting from the collective excitations of sea quarks for example are 
unaccounted for. One way of incorporating these additional ``sea quark'' effects without 
spoiling the agreement with the observed baryon spectra is to consider that
constituent quarks are effective degrees of freedom with some spatial extension.

Assuming that the up and down quarks are indistinguishable, a charge form factor 
$f(Q^2)$ of the Dirac type could  be appended to the charge- and current-density
operators  (both one- and two-body ones). A rather good agreement with data can be
obtained for $G_M^{p,n}$ at $Q^2> 0.5$ (GeV/$c$)$^2$ using a simple dipole
form factor 
\be
f(Q^2) = {1\over [1 + a Q^2]^2}
\label{eq:dipole}
\ee 
common to all quarks. Once constituent quarks are treated as extended
objects, it is not unreasonable to introduce an anomalous magnetic moment
$\kappa$. Thus, besides a dipole form for
$f(Q^2)$, the following form for $g(Q^2)$ has been considered:
\be
g(Q^2) = f(Q^2) + \kappa {1\over [1 + b Q^2]^3} 
\label{eq:cappa}
\ee
for the magnetic current-density operator.
The actual value of $\kappa$ has been fixed in order to obtain the
experimental value of the proton magnetic moment. For a quark mass $m=340$ 
one obtains $\kappa=0.379$. Correspondingly, the neutron
magnetic moment turns out to be $-2.073$ n.m. in good agreement with
experiment. The other two parameters $a$ and $b$ in eqs. (\ref{eq:dipole}) and
(\ref{eq:cappa}) are then fixed by fitting the $Q^2$ dependence of $G_M^p$. The
resulting value for the quark charge radius is: $r_c = 0.7$ fm. It is 
worth
noting that the extracted value of the quark charge radius is fairly
close to the value predicted by the Vector Meson Dominance model. Without
any additional free parameter one can then calculate the other nucleon 
form factors. The
results are shown in figs.~1 and~2 by the solid lines. 
A satisfactory agreement is obtained for both $G_E^p$ and $G_M^p$.

\section{Conclusions}

A completely consistent calculation of the nucleon elastic electromagnetic 
form factors has been performed within the extended GBE constituent quark 
model~\cite{gloz4}. We find that the two-body currents derived from the 
complete pseudoscalar, vector and scalar meson-exchange potentials by 
means of the continuity equation give rise to significant contributions to 
the proton and neutron magnetic moments, improving the agreement with the 
experimental values. 
 
However, both the electric and magnetic form factors predicted by the model 
overestimate
the observed nucleon form factors for $Q^2 > 0$ reflecting thus the 
inadequacy of the
assumption of pointlike constituent quarks. A satisfactory agreement is obtained when
treating the constituent quarks as
extended particles with anomalous magnetic moment using suitable Dirac- and Pauli-type
form factors. The resulting quark charge radius is consistent with
the prediction of the Vector Meson Dominance model.

\section*{Acknowledgements}

This work was partly performed under the TMR contract ERB FMRX-CT-96-0008.

\clearpage 

\begin{figure}[h]

\psfig{figure=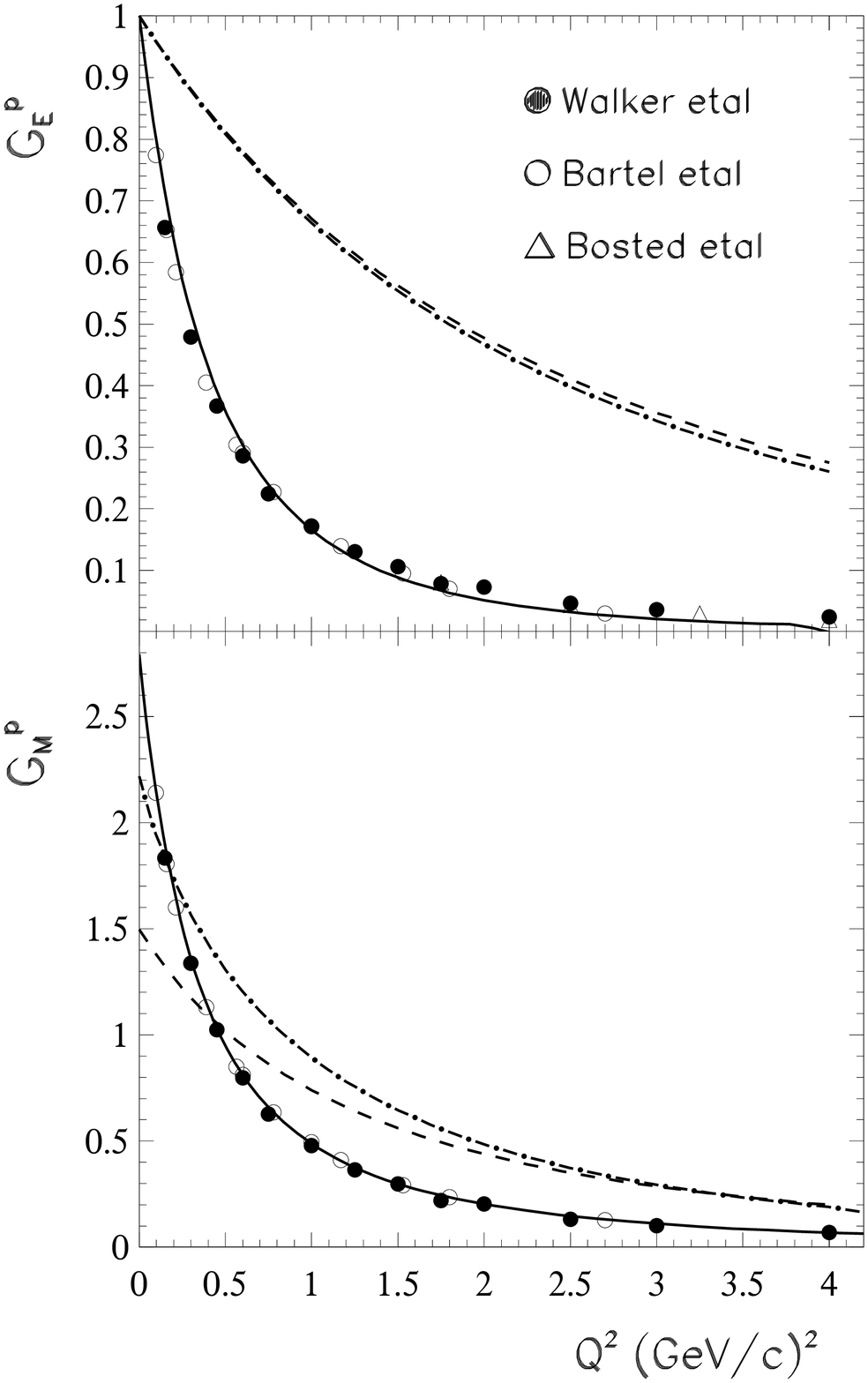,height=5in}

\caption{The electric ($G_E^p$) and magnetic ($G_M^p$) form factors  of 
the proton as a function of the four-momentum squared $Q^2$. 
The dashed, dot-dashed and solid lines refer to the results of the GBE 
CQM, extended GBE CQM and extended GBE CQM with quark form factors, 
respectively. Experimental points are from 
ref.~\protect\cite{Walker} (solid circles), ref.~\protect\cite{Bartel} 
(open circles) and ref.~\protect\cite{Bosted} (triangles).}

\end{figure}

\clearpage

\begin{figure}[h]

\psfig{figure=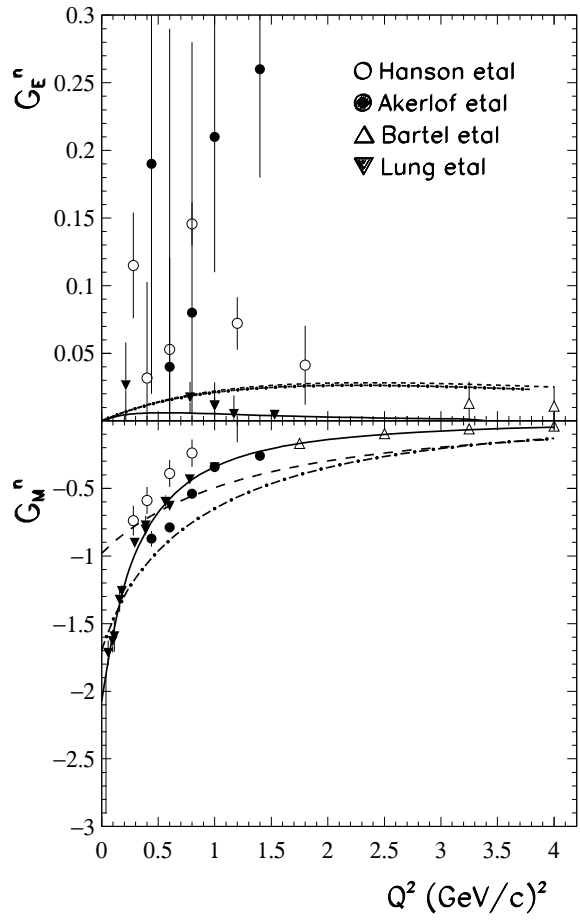,height=5in}

\caption{The same as in fig.1 but for the neutron. Experimental points are from ref.~\protect{\cite{Hanson}} (open circles), 
ref.~\protect\cite{Akerlof} (solid circles), 
ref.~\protect\cite{Bartel} (open triangles) and 
ref.~\protect\cite{Lung} (solid triangles).}

\end{figure}


\section*{References}
\begin{thebibliography}{99}

\bibitem{DeRujula}A.~De R\'ujula, H.~Georgi and S.~L.~Glashow,
\Journal{\PRD}{12}{147}{1975}.

\bibitem{IK} N.~Isgur and G.~Karl, \Journal{\PRD}{18}{4187}{1978}.

\bibitem{Giannini}M.M.~Giannini,\Journal{\RPP}{54}{453}{1990}.

\bibitem{Karl} G.~Karl, \Journal{\JMP}{1}{491}{1992}


\bibitem{Warns}M.~Warns, H.~Schr\"oder, W.~Pfeil and H.~Rollnik,
\Journal{\ZPC}{45}{613}{1990}.

\bibitem{Capstick}S.~Capstick, \Journal{\PRD}{46}{2864}{1992}.

\bibitem{Cardarelli}F.~Cardarelli, E.~Pace, G.~Salm\`e and S.~Simula,
\Journal{\PLB}{357}{267}{1995}.


\bibitem{Cotanch}A.~Szczepaniak, C.-R.~Ji and S.R.~Cotanch, \Journal{\PRC}{52}{2738}
{1995}.

\bibitem{Obukhovsky}I.T.~Obukhovsky and A.M.~Kusainov, \Journal{\PLB}{238}{142}{1990}.

\bibitem{BHF}A.~Buchmann, E.~Hernandez and A.~Faessler, \Journal{\PRC}{55}{448}{1997}.

\bibitem{valcarce}A.~Valcarce, P.~Gonz\'ales, F.~Fern\'andez and V.~Vento, 
\Journal{\PLB}{367}{35}{1996}.

\bibitem{fabre}Z.~Dziembowski, M.~Fabre de la Ripelle and G.A.~Miller,
\Journal{\PRC}{53}{R2038}{1996}.

\bibitem{Shen}P.N.~Shen, Y.B.~Dong, Z.Y.~Zhang, Y.W.~Yu and T.-S.H.~Lee, \Journal{\PRC}
{55}{2024}{1997}.

\bibitem{Manohar}A.~Manohar and H.~Georgi, \Journal{\NPB}{234}{189}{1984}.

\bibitem{gloz1}L. Ya.~Glozman and D.~O.~Riska, \Journal{\PP}{268}{1}{1996}.

\bibitem{gloz2}L.~Ya.~Glozman, Z.~Papp, W.~Plessas, K.~Varga and R.~F.~Wagenbrunn, 
\Journal{\PRC}{57}{3406}{1998}.


\bibitem{gloz3}L.~Ya.~Glozman, W.~Plessas, K.~Varga and R.~F.~Wagenbrunn,
\Journal{\PRD}{58}{094030}{1998}.

\bibitem{boffi1} S.~Boffi, P.~Demetriou, M.~Radici and
R.~F.~Wagenbrunn, \Journal{\PRC}{60}{025206}{1999}.

\bibitem{Capuzzi}S.~Boffi, F.~Capuzzi, P.~Demetriou and M.~Radici,
\Journal{\NPA}{637}{585}{1998}.

\bibitem{gloz4}W.~Plessas, L.~Ya.~Glozman, K.~Varga and R.~F.~Wagenbrunn, in
these Proceedings.


\bibitem{Walker}
R.C.~Walker et al., \Journal{\PLB}{224}{353}{1989}.

\bibitem{Bartel}W.~Bartel et al., \Journal{\NPB}{58}{429}{1973}.

\bibitem{Bosted}
P.~Bosted et al., \Journal{\PRL}{68}{3841}{1992}.


\bibitem{Hanson}
K.M.~Hanson et al., \Journal{\PRD}{8}{753}{1973}.

\bibitem{Akerlof}C.W.~Akerlof et al., \Journal{\PR}{135}{B810}{1964}.

\bibitem{Lung}A.~Lung et al., \Journal{\PRL}{70}{718}{1993}.


\end{thebibliography}
\end{document}